\documentstyle[a4,12pt]{article}
\begin{document}
\begin{flushright}
Liverpool Preprint: LTH 346\\
TFT Preprint: HU-TFT-95-21\\
hep-lat/9503018\\
14th March, 1995\\
\end{flushright}
\vspace{5mm}

\begin{center}

{\Large\bf Cooling and the $SU(2)$ instanton vacuum}\\[1.5cm]

{\bf C. Michael} \\
DAMTP, University of Liverpool, UK\\
{\bf  P.S. Spencer}\\
TFT, University of Helsinki, Finland\\
\end{center}

\begin{abstract}
We present results of an investigation into the nature of instantons
in 4-dimensional pure gauge lattice $SU(2)$\ obtained from configurations
which have been cooled using an under-relaxed cooling algorithm. We
discuss ways of calibrating the cooling and the effects of different
degrees of cooling, and compare our data for the shapes, sizes and
locations of instantons with continuum results.  In this paper we
extend the ideas and techniques developed by us for use in $O(3)$,
and compare the results with those obtained by other groups.
\end{abstract}

%
%

\section{Introduction}\label{s:intro}
Much work has been done in simulating the $SU(2)$ vacuum, with the aim of
learning more about the vacuum structure of QCD, and a lot of interest
has been shown in the existence of instantons in this vacuum. Early
studies into the cooled $SU(2)$ vacuum were carried out
in~\cite{tepersu2}, where, working on a lattice with `zero-derivative'
boundary conditions rather than the more usual periodic boundaries,
Teper investigated the role of instantons by locally minimizing the
action (ie. cooling) and concluded that
the lattice $SU(2)$ theory does indeed contain instantons, but that the
large, physically important ones will be absent in studies with
$\beta\leq 2.7$. At the same time, in~\cite{1stinst}, Ilgenfritz {\em
et al\/}, working on $6^4$ lattices at $\beta=2.1,2.2$, found
\mbox{(multi-)instanton} solutions from successive coolings of the
(equilibrated) gauge fields: \medskip
\begin{equation}\label{e:su2relax}
U_{x,\mu}\rightarrow U'_{x,\mu}=c\Sigma_{x,\mu}
\end{equation}
\noindent where $c$ is a normalisation constant to ensure
$U'_{x,\mu}\in SU(2)$ and $\Sigma$ is the sum of six staples (this
will be equivalent to taking $\alpha=0$ in eq.~\ref{e:su2ur}).  They
found that the action density is concentrated in distinct, separated
parts of the lattice~---~consistent with the picture of a dilute
instanton gas~---~and that these lattice instantons are meta-stable
under cooling of the gauge fields, being shrunk in size and finally
annihilated by prolonged cooling.  They also studied the eigenvalue
spectra of the fermion zero modes, and concluded that $SU(2)$ on the
lattice does indeed have an instanton vacuum.

In~\cite{polikarpov}, Polikarpov and Veselov, working on $8^4$--$12^4$
lattices at $\beta=2.3,2.35$ carried out a detailed study of the
cooling process and identified five distinct stages in the history of
a cooled configuration: (i)~freezing of quantum fluctuations and the
arising of quasi-classical configurations of instanton--anti-instanton
type; (ii)~smoothing of quantum fluctuations around inst\-antons and
anti-inst\-antons; (iii)~annihilation of instantons and
anti-instantons; (iv)~a long stage involving the existence of
classical solutions; (v)~a short stage involving the disappearance of
classical solutions.

The classification of these stages is, in their own words, somewhat
arbitrary, with an absence of definite borders between the
stages. They found that there were (at least) two characteristic
points on the cooling curve, however: around the middle of stage (iv)
they found a zero of $\Delta^2S(N)$ (corresponding to the second
derivative of $S$ with respect to cooling sweep $N$), and another
somewhere on the border between stages (ii) and (iii) at which
$\Delta^2S(N)\approx 0$.  They calibrated their cooling process by
stopping at the former point; they denoted a configuration in this
state as an `instanton vacuum', and identified the latter point as the
beginning of the instanton--anti-instanton annihilation, denoted as
the `instanton--anti-instanton vacuum'.

\medskip

This paper is organised as follows: in section~\ref{s:su2conti} we
outline the formalism of continuum $SU(2)$, and briefly discuss the
instanton solution found by Belavin {\em et al\/} in~\cite{belavin};
in section~\ref{s:su2latti} we consider how to construct an $SU(2)$
instanton on a periodic lattice; in section~\ref{s:calib} we discuss
calibrating the cooling we use, with a first calibration attempt given
in subsection~\ref{s:calib1} and a better method given in
subsection~\ref{s:calib2}; in section~\ref{s:dist} we detail the means
of locating and sizing the instantons, compare their shapes with the
form of isolated, continuum objects and calculate a size distribution
for them; in section~\ref{s:seps} we consider inter-instanton
separations and compare our results with
refs.~\cite{forster,palmer}. Our conclusions are given in
section~\ref{s:su2conc}. The figures referred to in the text are
collected together at the end of the paper. Preliminary results from
this work were presented at Lattice 94~\cite{us:lat94}.

%
%

\section{Instantons in $SU(2)$}\label{s:su2conti}

As in 2-dimensional $O(3)$, 4-dimensional $SU(2)$ has instanton
solutions\protect\footnote[1]{This section follows the
discussion of Yang-Mills instantons given
in~\cite{chengli}.} due, in this case, to the
homotopy class of mappings from $S_3\rightarrow S_3$, ie from the
points on a four-dimensional sphere parameterised by three angles, to
the elements of $SU(2)$, which is also determined by three parameters. In
other words, the $SU(2)$ group manifold is topologically equivalent to
$S_3$: any $SU(2)$ group element can be written in terms of the Pauli
matrices as
\begin{equation}\label{e:su2u_2}
U=u_4{\bf 1}+\mbox{\rm i} \mbox{\boldmath $ u\cdot\sigma$ }\ ;\
u^2_4+\mbox{\boldmath $u$}^2=1\ ,
\ \mbox{\boldmath $\sigma$}=(\sigma_1,\sigma_2,\sigma_3)
\end{equation}
\noindent with ${\bf 1}$ the $2\times 2$ unit matrix, and
$u=(u_4,\mbox{\boldmath $u$})$ clearly defining a point on $S_3$, the sphere in
4-dimensional Euclidean space. Mappings in this case are characterised
by the topological charge $Q^T$ defined in terms of the divergence of
a (gauge-dependent) current:
\begin{eqnarray}\label{e:su2q}
Q^T&=&\frac{1}{16\pi^2}\int d^4x\;\mbox{\rm tr}\; F_{\mu\nu}\tilde F_{\mu\nu}\\
\partial_{\mu}K_{\mu}&=&2\mbox{\rm tr}\; (F_{\mu\nu}\tilde F_{\mu\nu})\\
K_{\alpha}&=&4\varepsilon_{\alpha\beta\gamma\delta}\mbox{\rm tr}\;
(A_{\beta}\partial_{\gamma}A_{\delta}+\frac{2}{3}A_{\beta}A_{\gamma}A_{\delta})
\end{eqnarray}
\noindent with the gauge fields and field strength tensor scaled by $\mbox{\rm
i}/g$
and defined by
\begin{equation}\label{e:fmunu}
F_{\mu\nu}=\partial_{\mu}A_{\nu}-\partial_{\nu}A_{\mu}+[A_{\mu},A_{\nu}]
\end{equation}
\noindent and the dual of $F$ given by $\tilde
F_{\mu\nu}=\frac{1}{2}\varepsilon_{\mu\nu\lambda\rho}F_{\lambda\rho}$.
Under a gauge transformation $T(x_{\mu})$ we have
\begin{equation}\label{e:su2gt}
A_{\mu}\rightarrow T^{\dagger}A_{\mu}T+T^{\dagger}\partial_{\mu}T\ .
\end{equation}
To find the instanton solutions we look for finite-action solutions to
classical Euclidean Yang-Mills theory. In order to have the action
\begin{equation}\label{e:su2s}
S=\frac{1}{2g^2}\int d^4x\;\mbox{\rm tr}\; F_{\mu\nu}F_{\mu\nu}
\end{equation}
\noindent finite we require
$F_{\mu\nu}(x)\stackrel{|x|\rightarrow\infty}{\rightarrow}0$ which we
can take to mean
$A_{\mu}(x)\stackrel{|x|\rightarrow\infty}{\rightarrow}0$.  However,
in practice this is too restrictive, and eq.~\ref{e:su2gt} means
we only need
\begin{equation}\label{e:su2bc}
A_{\mu}(x)\rightarrow T^{\dagger}\partial_{\mu}T\ {\rm as}
\ |x|\rightarrow\infty
\end{equation}
\noindent which is obtained from $A_{\mu}=0$ by a gauge transformation. To
find the fields satisfying this boundary condition we make use of the
following positivity relation
\begin{equation}\label{su2+ve}
\mbox{\rm tr}\; \int d^4x\;(F_{\mu\nu}\pm\tilde F_{\mu\nu})^2\geq 0\ .
\end{equation}
\noindent As
\begin{equation}\label{e:su2f2}
(F_{\mu\nu}\pm\tilde
F_{\mu\nu})^2=2(F_{\mu\nu}F_{\mu\nu}\pm F_{\mu\nu}\tilde F_{\mu\nu})
\end{equation}
\noindent we have
\begin{equation}\label{e:su2ineq}
\mbox{\rm tr}\; \int d^4x\;F_{\mu\nu}F_{\mu\nu}\geq\left|\;\mbox{\rm tr}\;\int
d^4x\;F_{\mu\nu}\tilde
F_{\mu\nu}\right|=16\pi^2Q
\end{equation}
\noindent using the definition for $Q$ given in eq.~\ref{e:su2q}. From this
we have
\begin{equation}\label{e:su2slb}
S\geq\frac{8\pi^2}{g^2}Q
\end{equation}
\noindent whence we see that the action for a single instanton is
$S_I=8\pi^2/g^2$ (cf. the result for $O(3)$: $S_I=4\pi/g^2$). Clearly,
then, the action is minimised for the (anti-)self dual solutions
\mbox{$F_{\mu\nu}=\pm\tilde F_{\mu\nu}$.}

In looking for the instanton solution, Belavin {\em et al\/}~\cite{belavin}
considered first $O(4)$ gauge theory, isomorphic to
$SU(2)\times SU(2)$ with one $SU(2)$ identified with the internal symmetry and
the other with the three-sphere at $|x|\rightarrow\infty$. The resulting
gauge transformation has the form
\begin{equation}\label{e:su2igt}
T(x)=\frac{1}{|x|}(x_4+\mbox{\rm i} \mbox{\boldmath $x\cdot\sigma$})\
{\rm with}
\ |x|^2=x_4^2+\mbox{\boldmath $x$}^2
\end{equation}
\noindent giving rise to a gauge field
\begin{equation}\label{e:su2ig}
A_{\mu}(x)=\frac{x^2}{\rho^2+x^2}T^{\dagger}(x)\partial_{\mu}T(x)
\end{equation}
\noindent where $\rho$ is the instanton size, and $x=0$ its
centre. For $x^2\gg\rho^2$, $A_{\mu}\rightarrow T^{\dagger}\partial_{\mu}T$
as required by eq.~\ref{e:su2bc}.  We can write the space and time
components of $A_{\mu}$ explicitly as
\begin{eqnarray}\label{e:su2igc}
A_4(x)&=&\frac{-\mbox{\rm i}\mbox{\boldmath $x$}\cdot
\mbox{\boldmath $\sigma$}}{ x^2+\rho^2}\\
\mbox{\boldmath $A$}(x)&=&\frac{\mbox{\rm i}(x_4\mbox{\boldmath
$\sigma$}+
\mbox{\boldmath $\sigma$}\times\mbox{\boldmath $x$})}
{ x^2+\rho^2}\ .
\end{eqnarray}

\subsection{Instanton--anti-instanton interactions}\label{s:int}
 Thus far we have only considered the single-instanton solution;
multi-instanton configurations can be generated by simply multiplying
instanton fields with their centres at different locations~\cite{smit},
and calculations involving mixed instanton--anti-instanton
configurations are non-trivial. In~\cite{forster}, F\"or\-ster
considered an inst\-an\-ton--anti\--inst\-an\-ton pair at large
separation, taking an instanton at $x=0$, given by eq.~\ref{e:su2ig},
and an anti-instanton at $x=C$, given by:
 \begin{equation}\label{e:su2aig}
A'_{\mu}(x)=\frac{(x-C)^2}{\lambda^2+(x-C)^2}T(x-C)
\partial_{\mu}T^{\dagger}(x-C)
\end{equation}
\noindent with $\lambda$ the size parameter for the
anti-instanton. By performing a conformal mapping, the point $x=C$ is
sent to infinity and the local behaviour at $x=C$ implies
the interaction takes the form of a kink and an anti-kink separated by
a distance $-\log(\rho\lambda)$ giving an interaction term:
\begin{equation}\label{e:su2int}
S_{\rm int}= -48\frac{\pi^2}{g^2}\frac{\rho^2\lambda^2}{(x_1-x_2)^4}
\end{equation}
\noindent with $g^2$ the coupling occuring in eq.~\ref{e:su2s} above,
implying an attraction between unlike objects. In~\cite{palmer},
Palmer and Pinsky, investigating this interaction and the distinction
between dilute and dense phases of the instanton gas, considered the
case of an instanton--anti-instanton pair separated by the size of one
member, with the size of the other free. They found that there was a
minimum finite separation:
\begin{equation}\label{e:su2rmin}
c=\rho=2\lambda
\end{equation}
\noindent at which point the pair melts into a pure gauge with zero
action. This will be relevant when we investigate the separations of
objects in section~\ref{s:seps} below.

%
%

\section{$SU(2)$ instantons on the lattice\label{s:su2latti}}
 We seek to extend the successful calibrated cooling
method~\cite{us:cool,us:dist}  from $O(3)$  to $SU(2)$.
As before, we use an under-relaxed cool, in this case given by
 \medskip
\begin{equation}\label{e:su2ur}
U'_{x,\mu}=\Sigma_{x,\mu}+\alpha U_{x,\mu}
\end{equation}
 \noindent where $\Sigma_{x,\mu}$ is the sum of the six `staples' around
the link $U_{x,\mu}$, and $U'_{x,\mu}$ is subsequently normalised to lie
in $SU(2)$. Following the results of~\cite{us:cool,us:dist} we take
$\alpha=2$.

In order to calibrate the cooling, we want to generate a gauge
configuration by using a suitable lattice version of
eq.~\ref{e:su2ig}, as we intend to calibrate the cooling by examining
its effects on a particular known configuration. However, this formula
is explicitly incompatible with periodic boundary
conditions\footnote[1]{We are greatly indebted to Mike Teper for
discussion on this subject.}, as we require the $A_{\mu}$ fields to
vanish at large distances from the instanton centre\footnote[2]{The
concept of ``large distance'' on a periodic lattice may seem a strange
one. What we need, for a moderately sized instanton sited at the centre
of the lattice, is that the $A_{\mu}$ fields approach zero at the
lattice boundary.}. In order to achieve this we must construct the
instanton as follows~\cite{mjtinst}: first construct the $A_{\mu}(x)$ in
$R^4$ as above. Next construct the link variables $U_{x,\mu}$ from these
continuum fields. As stated above, we require an instanton solution to
be self-dual, and for the lattice solution this relation is only
approximate. If we take as the link fields
 \begin{equation}\label{e:su2iU}
U_{x,\mu}=\exp(\mbox{\rm i}[A_{\mu}(x)+A_{\mu}(x+\mu)]/2)
\end{equation}
\noindent constructing these on an infinite lattice, then the deviation
from self-duality is ${\cal O}(\rho^{-2})$~\cite{mjtinst}. Next we
perform a singular gauge transformation {\em on the lattice\/} using
eq.~\ref{e:su2igt}, with the singularity `off-site' both here and in the
construction of the $A_{x,\mu}$ fields:
\begin{equation}\label{e:su2sgt}
U_{x,\mu}=T_{x}U_{x,\mu}T^{\dagger}_{x+\mu}
\end{equation}
\noindent (where $T_x$ denotes the gauge transformation on the site $x$
of the infinite lattice) and finally we impose periodicity on the
solution by enforcing
\begin{equation}\label{e:su2per}
U_{\mu}(x_i=-L/2)\equiv U_{\mu}(x_i=+L/2)
\end{equation}
\noindent for each $x_i$ and each $\mu$ for a $L^4$ lattice.

In constructing this solution we must also take into account the fact
that eq.~\ref{e:su2ig} gives the instanton in terms of the
$A_{\mu}(x)$ fields, while on the lattice we need to use the
$U_{x,\mu}$ link fields, and so need to rewrite the instanton solution
in terms of these link variables. This is straightforward as
\begin{equation}\label{e:infsu2}
\exp(\mbox{\rm i}\theta \hat{\mbox{\boldmath $x$}}\cdot
\mbox{\boldmath $\sigma$})=\cos\theta+\mbox{\rm i}
(\hat{\mbox{\boldmath $x$}}\cdot\mbox{\boldmath $\sigma$})\sin\theta
\end{equation}
\noindent and we can rewrite eq.~\ref{e:su2ig} in this form. The
individual components are, explicitly:
\begin{eqnarray}\label{e:su2iu}
\mbox{\rm i} A_4(x)&=&\frac{1}{x^2+\rho^2} (\mbox{\boldmath
$\sigma$}\cdot(x_1,x_2,x_3))\\
\mbox{\rm i} A_1(x)&=&\frac{-1}{x^2+\rho^2}(\mbox{\boldmath
$\sigma$}\cdot(x_4,x_3,-x_2))\\
\mbox{\rm i} A_2(x)&=&\frac{-1}{x^2+\rho^2}(\mbox{\boldmath
$sigma$}\cdot(-x_3,x_4,x_1))\\
\mbox{\rm i} A_3(x)&=&\frac{-1}{x^2+\rho^2}(\mbox{\boldmath
$sigma$}\cdot(x_2,-x_1,x_4))\ .
\end{eqnarray}

As in $O(3)$, this field configuration is metastable (with the Wilson
gauge action we are using) and will be
annihilated by prolonged cooling. (See the discussion
in~\cite{us:cool,us:dist}). For instantons where $\rho$ is not small
compared to $L$ there will be significant discontinuities across the
lattice boundary, which can be removed by a moderate amount of
cooling.

%
%

\section{Calibrating the under-relaxed cooling.}\label{s:calib}
\begin{figure}[th]
\vspace{3 in}
\includegraphics{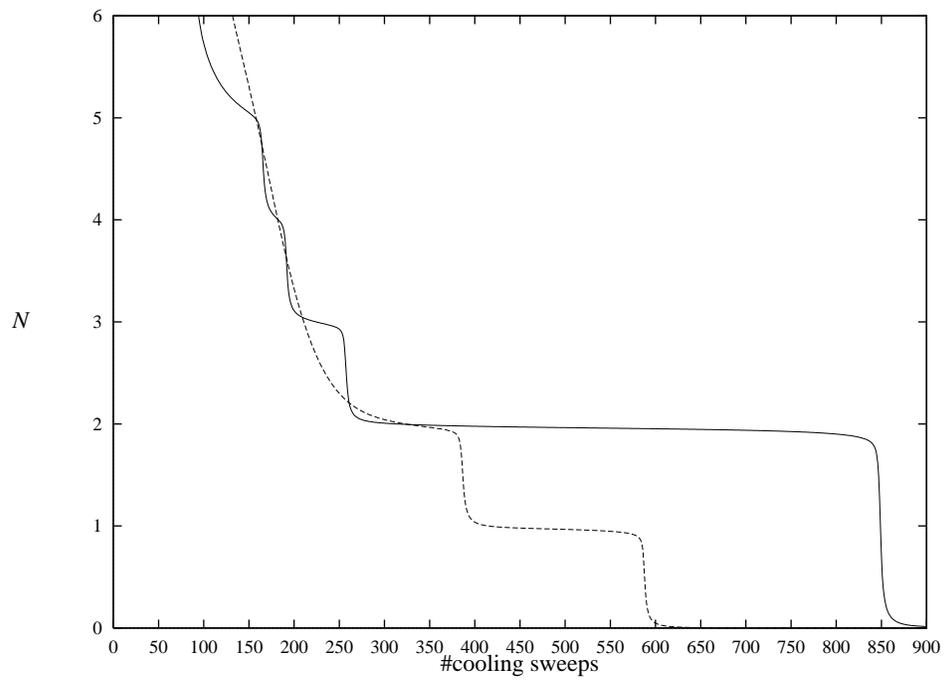}
\vspace{.5 in}
\caption{$N$ ($S/S_I$) for two example $SU(2)$ configurations as they
are cooled with $\alpha=2$.\label{f:su2cool}}
\end{figure}

We performed our simulations on a CRAY Y-MP, using $16^4$ and
$24^4$ lattices at $\beta=2.4,\,2.5$ respectively.  These correspond
to lattice spacings of approximately $0.12{\rm fm}$ and $0.08{\rm
fm}$~\cite{green}.  Each of our configurations is generated from 500
updates with 1 heatbath update followed by 3 over-relaxation
updates, after several thousand thermalisation sweeps.  In
fig.~\ref{f:su2cool} we present the cooling history of two example
$SU(2)$ gauge configurations, calculated on the smaller lattice.

\subsection{A first attempt at calibration}\label{s:calib1}
For $O(3)$, we concluded~\cite{us:dist} that an optimum cooling would be
that required to annihilate an instanton of size $\rho={\cal O}(2a)$
where $a$ is the lattice spacing.  In table~\ref{t:su2kill} there is
shown the number of cooling sweeps at $\alpha=2$ (which we had decided
was a good compromise between cooling too severely and making too many
cooling sweeps) needed to
remove $SU(2)$ instantons of size $\rho=a,3a/2,2a,4a$ generated from
the prescription above.  It transpires that these are much more stable
under cooling than their $O(3)$ counterparts.

\begin{table}[h]
\begin{center}
\begin{tabular}{ccccc}\hline
$\rho$           & $a$ & $3a/2$ & $2a$ & $4a$ \\ \hline
$N_{\rm sweeps}$ & 495 & 503    & 510  & 536  \\ \hline
\end{tabular}
\caption{\label{t:su2kill} The number of sweeps $N_{\rm sweeps}$ at
$\alpha=2$ required to annihilate (to reduce $S/S_I$ to~$<0.1$) a single
$SU(2)$ instanton of size $\rho$. Calculated on a $16^4$ lattice at
$\beta=2.4$, with the instanton positioned off-site in the centre of
the lattice.}
\end{center}
\end{table}

This much greater stability presents us with something of a dilemma:
were we to na\"{\i}vely proceed and take the value for $\rho=2a$ in
table~\ref{t:su2kill} as a calibrator, then the gauge configurations
we would have after cooling would be well past the first three stages
defined by Polikarpov and Veselov above~---~which is where we wish to
examine the vacuum, as we wish to study its structure in the presence
of \mbox{(anti-)}instantons~---~and in some cases will have been
cooled so far as to entirely remove all vacuum structure. There are
two possible candidates from the cooling curve of annihilation of a
single instanton for the level of cooling to use. The first is the
cooling required to remove the undesired boundary effects, while the
second is the amount required to then annihilate the instanton, once
the boundary discontinuities have been eliminated. However, neither of
these methods of calibration is really satisfactory~---~the former is
too unphysical a process, as it removes what is essentially a lattice
artefact that has nothing to do with the underlying phyics, and given
the size of the lattice and the value of $\beta$ we use, the latter
seems excessive when compared to the results presented
in~\cite{us:dist}. In light of this we propose a different means of
calibration.

\subsection{A second calibration method}\label{s:calib2}
As we wish to model the vacuum as quantum fluctuations around
classical configurations, we first generate a classical configuration:
the single-instanton solution as presented above.  We then cool this
by enough sweeps to remove any boundary discontinuities. Next we
perform a few heatbath updates at the $\beta$ value under study to
introduce some quantum fluctuations and then cool it until we regain
the original configuration, ie until $S=S_I\pm 10\%$.

Quite how many heatbath sweeps to perform is open to question: we seek
to reach a state in some respects similar to the uncooled
configurations, but not too far away from the original instanton~---~we
do not want to introduce any more structure than was already present.
We found that, at $\beta=2.4$ on a $16^4$ lattice, 10 heatbath sweeps
was optimum for introducing the required quantum fluctuations without
overly changing the underlying structure, whereas for $\beta=2.5$ on a
$24^4$ lattice, 10 was too few and, at this higher $\beta$ value, we
found 20 were required. As a check of this we found 20 too many for the
lower $\beta$ on the larger lattice, with new structure occuring in more
than half of the sample configurations we looked at; by comparison, for
$\beta=2.5$, 20 sweeps introduced new structure\footnote[1]{We ascertained
whether the configuration had been moved away from the single-instanton
solution by following the cooling history. Where new structure had been
introduced by the heatbath sweeps, `steps' appeared in the history, in
much the same way as they appear in the solid line in
fig.~\protect\ref{f:su2cool}.} in less than $5\%$ of the ensemble.

As a further check of the validity of the number of thermalisation
sweeps used, we looked at the value of $\langle{\rm tr}
U_{\Box}\rangle$ on the configuration. We found that with 10
heating sweeps on the $\beta=2.4,\,16^4$ lattice this was consistent
with established results, and this was also true for 20 sweeps on the
$\beta=2.5,\,24^4$ configurations, while half the number of sweeps in
each case gave a value that was too high. These two checks~---~not
introducing more structure, and heating until the correct value of
$\langle{\rm tr}U_{\Box}\rangle$ is attained at a given $\beta$
value~---~form respectively effective upper and lower bounds on the
number of heatbath sweeps to be performed.

The calibration results are given in table~\ref{t:qfswps} for initial
configurations each containing a single instanton of size
$\rho=a,2a,3a,4a$.

In~\cite{us:dist} we decided that a good check on whether the cooling
had taken the configuration to the same physical state was whether or
not there was a good agreement across $\beta$ of the value of $\langle
S/V_{\rm phys}S_I\rangle$, and found that the criteria we used for $O(3)$
led to such an agreement. For our $SU(2)$ data however, the agreement
appears to be less good, with $\langle S/V_{\rm
phys}S_I\rangle=0.000185(2)$ at $\beta=2.4$ and $\langle S/V_{\rm
phys}S_I\rangle=0.000249(3)$ at $\beta=2.5$.  While not equal, these
two numbers seemed sufficiently close for us to accept this
calibration method, although no doubt these values could be brought
closer by a certain amount of `fine-tuning'. On the basis of these
results we chose 72 sweeps to be the number required at $\beta=2.4$
with $\alpha=2$, and 154 sweeps at $\beta=2.5$, and accepted this
method as a calibration technique for under-relaxed cooling in $4d$
$SU(2)$.

\begin{table}[t]
\begin{center}
\begin{tabular}{cccccc}\hline
$\beta$ & $L$ & $N_{\rm heat}$ & $\rho$ & $N_{\rm cool}$ &
$\langle{\rm tr}U_{\Box}\rangle$\\
\hline\hline
$2.4$   & 16  & 10 & $a$      & 71(8)          &  0.6344(6) \\
        &     &    & $2a$     & 72(7)          &  0.6346(5) \\
        &     &    & $3a$     & 71(7)          &  0.6347(7) \\
        &     &    & $4a$     & 73(5)          &  0.6345(5) \\\hline
$2.5$   & 24  & 20 & $a$      & 152(12)        &  0.6529(3) \\
        &     &    & $2a$     & 155(8)         &  0.6530(3) \\
        &     &    & $3a$     & 154(11)        &  0.6530(3) \\
        &     &    & $4a$ & 154(13) & 0.6530(3) \\ \hline \end{tabular}

\caption{\label{t:qfswps} The number of sweeps $N_{\rm cool}$ at
$\alpha=2$ required to restore a configuration with a single lattice
instanton of size $\rho$ after $N_{\rm heat}$ heatbath updates, and the
value of the average plaquette after the heatbath sweeps.  The figures
in parentheses are one standard deviation, the data are obtained from 50
configurations in each case.  The accepted values of $\langle{\rm
tr}U_{\Box}\rangle$ are $0.63058(2) [\beta=2.4]$ and
$0.65235(1) \mbox{$[\beta=2.5]$}$~\protect\cite{karsch}.}
\end{center}
\end{table}

%
%

\section{Instanton size, shape and distributions}\label{s:dist}
In~\cite{us:dist} we found we could obtain good results for the sizes
and locations of instantons using $S(x)$, although in principle
$Q(x)$ contains more information. This was because in investigating
the size distribution we were not concerned with the sign of the
object. When we came to look at the inter-instanton separations, we
still used $S(x)$ for calculating the locations and obtained
information on the sign from $Q(x)$.

For $SU(2)$, we used the `plaquette' method~\cite{campopl} to calculate
$Q(x)$ on our  cooled configurations:
\begin{equation}\label{e:plaqq}
Q_L(x)=\frac{1}{64\pi^2}\left(\mbox{\rm tr}\;(\Pi_{x,12}\Pi_{x,34})+
\mbox{\rm tr}\;(\Pi_{x,13}\Pi_{x,42})+\mbox{\rm
tr}\;(\Pi_{x,14}\Pi_{x,23})\right)
\end{equation}
\noindent where $\Pi_{x,ij}$ is the orientated path combination at $x$ in
the $ij$ plane, consisting of the sum of the four clover leaves of
$(U_{\Box}-U^{\dagger}_{\Box})/2$.

\begin{figure}[bh]
\vspace{2.5 in}
\includegraphics{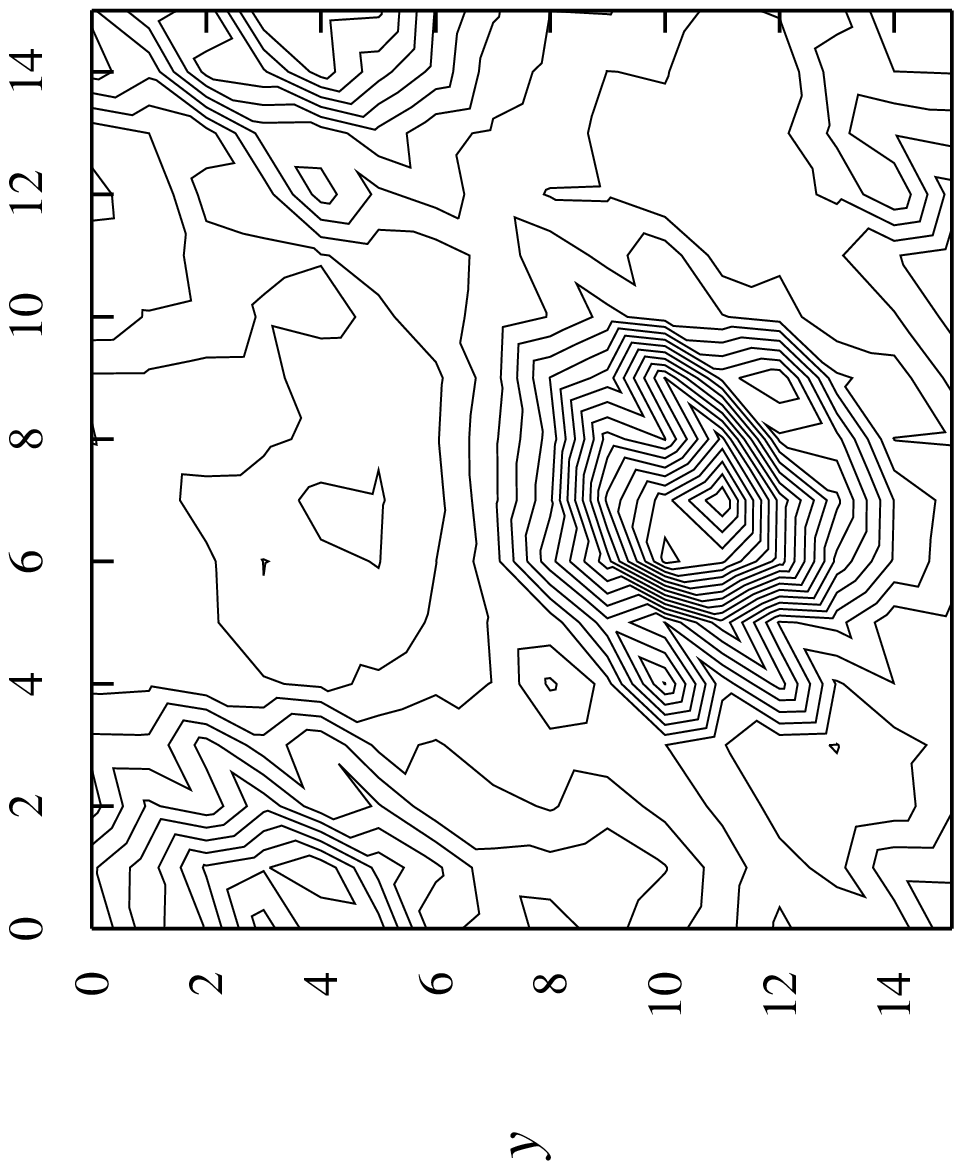}
\includegraphics{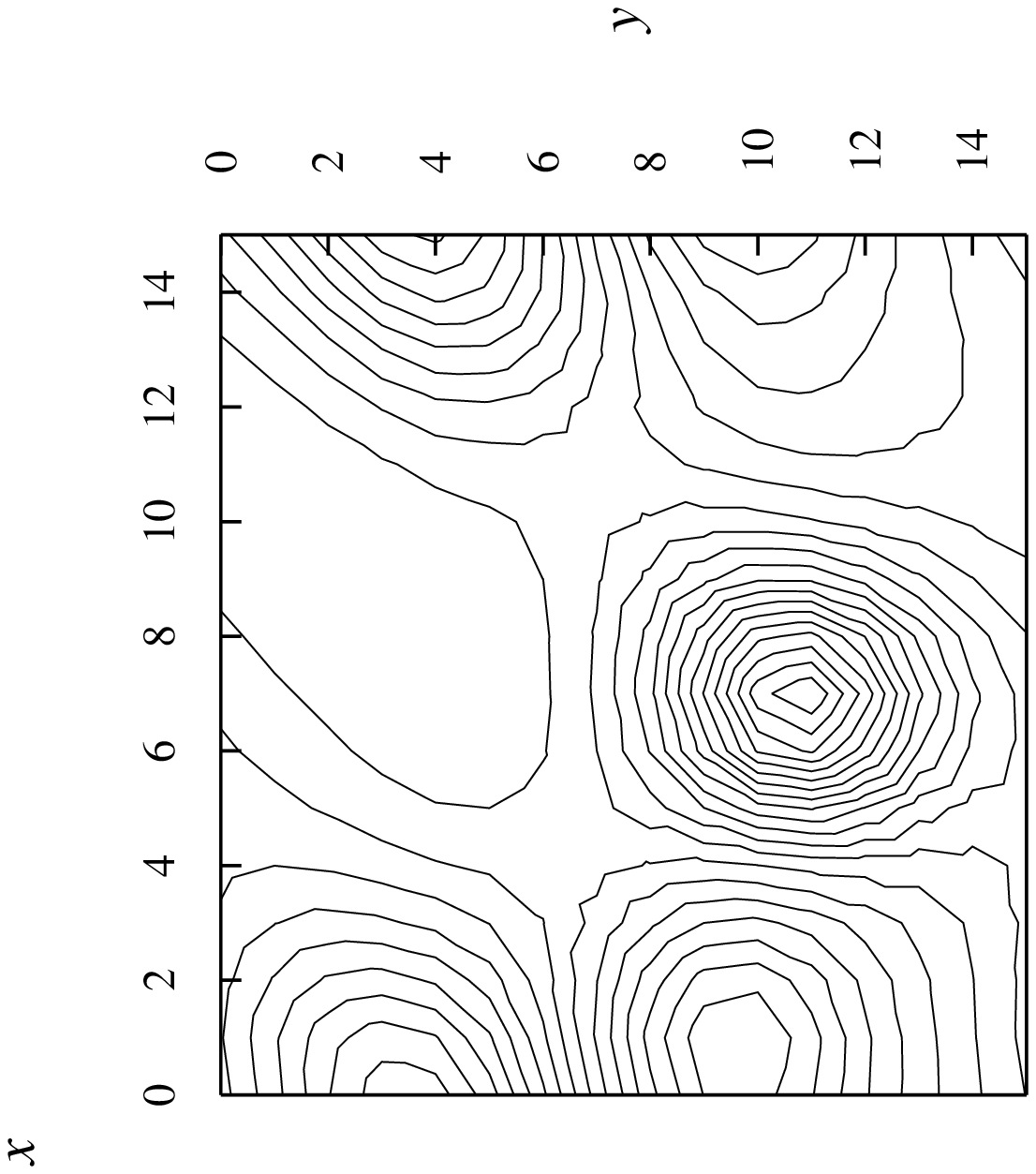}
\vspace{.5 in}
\caption{\label{f:cont}Contour plots for fig.~\protect\ref{f:svsqlink}.
$S(x,y,14,9)$ is the left plot and $Q(x,y,14,9)$ is the right plot.}
\end{figure}

\begin{figure}[p]
\vspace{6in}
\includegraphics{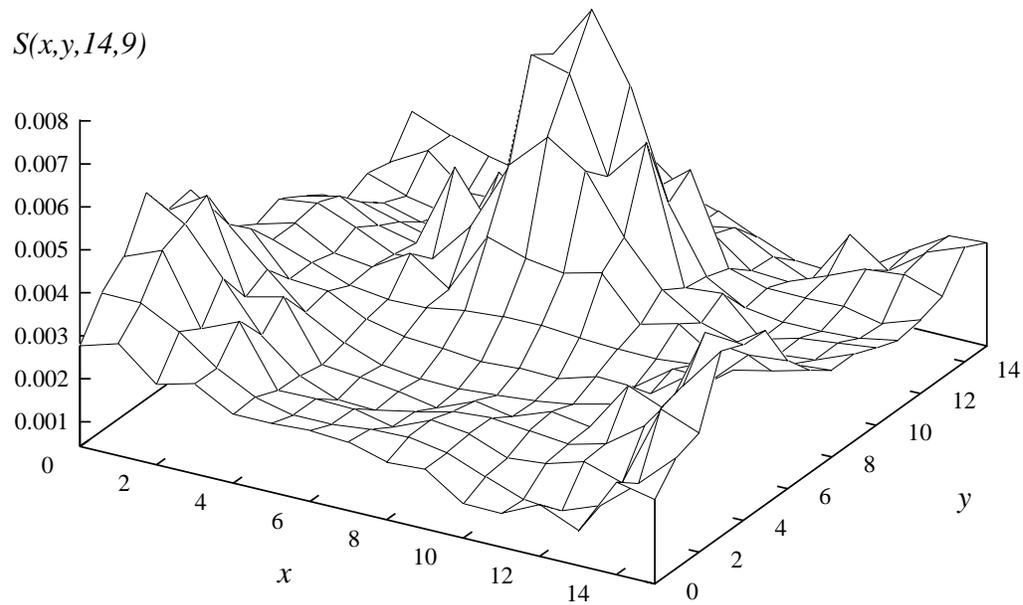}
\includegraphics{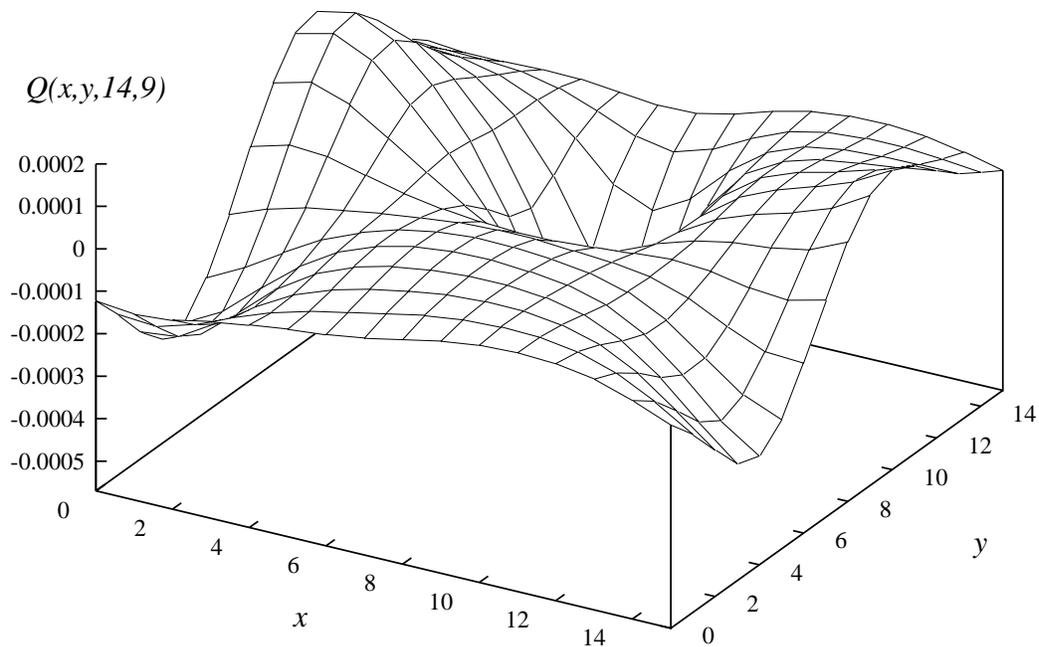}
\vspace{0.75in}
\caption{\label{f:svsqlink} The action and topological charge densities
for the $(x,y,14,9)$ plane from a sample configuration, generated at
$\beta=2.4$ on a $16^4$ lattice.}
\end{figure}

We choose the most local and symmetric definition for the action density
at a site. We used the average of contributions from all plaquettes
with one corner on the site $x$. This results in an action density
that matches $Q(x)$ very well (and in the continuum $S(x)\propto Q(x)$
around an isolated instanton), but is not as smooth as $Q(x)$, as can
be seen from figs.~\ref{f:cont} and~\ref{f:svsqlink}.  Since $S(x)$
not as smooth as $Q(x)$, on occasion the instanton location algorithm
will record more structure in a given area of the lattice than is
actually there and will ignore valid structure elsewhere when applied
to $S(x)$. For this reason we conducted our instanton search at
$\beta=2.4$ using both $S(x)$ and $Q(x)$.

\bigskip

\begin{figure}[th]
\begin{center}
\input{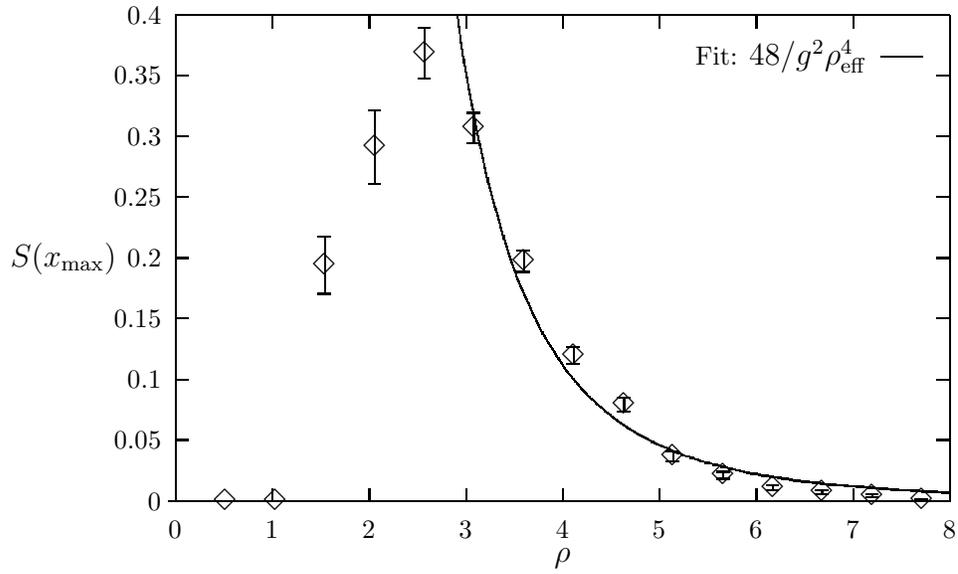}
\caption{\label{f:hvsws}Peak height against instanton size $\rho$ (in
units of the glueball mass) calculated
using $S(x)$ on a $16^4$ lattice at $\beta=2.4$.}
\end{center}
\end{figure}

\begin{figure}[bh]
\begin{center}
\input{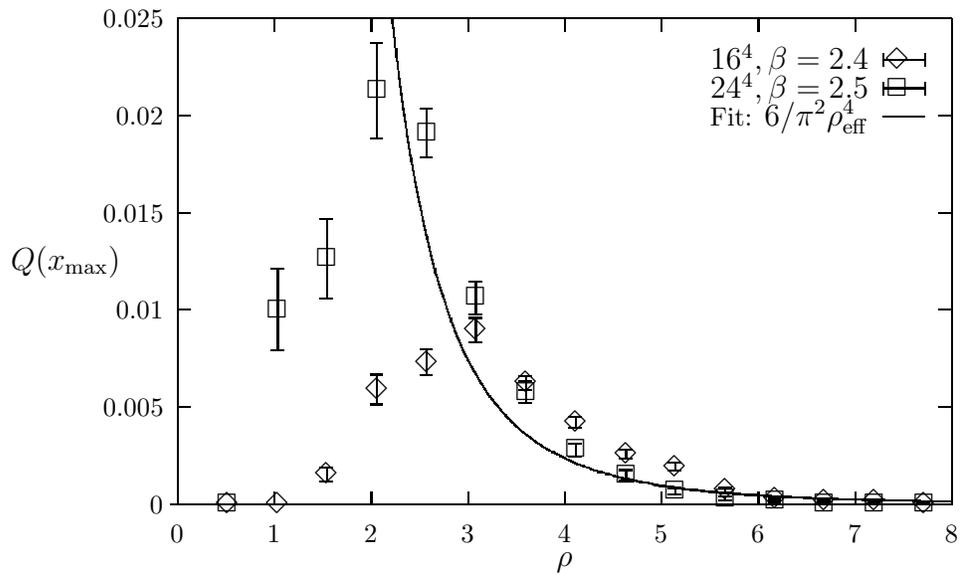}
\caption{\label{f:hvswq}Peak height against instanton size $\rho$ (in
units of the glueball mass) calculated
using $|Q(x)|$.}
\end{center}
\end{figure}

Here we look at connected 4-dimensional regions of the lattice. We use
the same method as in~\cite{us:dist} by looking for local maxima in
the density, and in this case use as a size parameter, which we denote
$\rho_L$, the 4th root of the hypervolume of the connected region of
sites for which the density is not less than half that at the peak.
In order to relate this $\rho_L$ back to the size parameter in
eq.~\ref{e:su2ig}, we define an effective radius (see
ref.~\cite{grandyl94} for a similar prescription) $\rho_{\rm eff}$:
\begin{equation}\label{e:rhoscale}
\rho_{\rm eff}^2=\frac{\sqrt{2}}{\pi(\sqrt[4]{2}-1)}\,\rho_L^2\ .
\end{equation}
\noindent This is the parameter plotted in
figs.~\ref{f:hvsws}--\ref{f:dist2} and quoted in
table~\ref{t:seps}. Eq.~\ref{e:su2slb} gives a lower bound on the
action, and so $S/S_I$ can be interpreted as an upper bound on the
number of objects in a given configuration, and we used this to
determine how many we looked for in each configuration..

To see how closely the individual objects resemble isolated continuum
instantons, we looked at the relation between the height at the peak
and the calculated size, as these have the continuum relations
\begin{equation}\label{e:su2svsr}
S_{\rm max}=\frac{48}{g^2}\frac{1}{\rho^4}
\end{equation}
\noindent and
\begin{equation}\label{e:su2qvsr}
Q_{\rm max}=\frac{2g^2}{16\pi^2}S_{\rm max}=\frac{6}{\pi^2\rho^4}
\end{equation}
\noindent for an isolated instanton. Our results are presented in
figs.~\ref{f:hvsws} and~\ref{f:hvswq}. All distances are in units of
the lightest glueball mass, $\xi=1/m_{0^+}$ with
$m_{0^+}(2.4)=0.94(8)$ and $m_{0^+}(2.5)=0.63(8)$, taking the values
for the lattice spacing $a$ at these values of $\beta$
from~\cite{green} and the mass of the $0^+$ glueball at $\beta=2.4$
from~\cite{su2mass}. The data was obtained from 261 configs at
$\beta=2.4$ and 250 at $\beta=2.5$.

As in~\cite{us:dist} we calculated the distribution of instanton
sizes. The data for $dN/Vd\rho$ are presented in
figs.~\ref{f:dist1}--\ref{f:dist2}. All sizes are again in units of
the correlation length, $\xi$. (It should be noted that the plot we
presented earlier in~\cite{us:lat94} used $\rho_L$, scaled by a
different factor from that above, rather than $\rho_{\rm eff}$ and so
contained a peak in a different position.)

In agreement with~\cite{1stinst,polikarpov,chu} we find a distribution
that grows with instanton size (unlike that presented for $O(3)$
in~\cite{us:dist}), although our distribution lacks the sharp
large-size cutoff they find; we attribute this to our means of
determining the size from connected hyper-volumes whereas
in~\cite{1stinst,polikarpov} they simply measured the size along the
lattice axes. In~\cite{chu}, Chu {\em et al\/} measured the size by
using the correlation function
\begin{equation}\label{e:qxqy}
f(x)=\sum_yQ(y)Q(x+y)
\end{equation}
\noindent and comparing this with a convolution of the analytic
expression:
\begin{equation}\label{e:qx}
Q_{\rho}(x)=\frac{6}{\pi^2\rho^4}\left(\frac{\rho^2}{x^2+\rho^2}\right)^4
\end{equation}
\noindent which they fitted to the lattice data for $f(x)$, obtained
by cooling for 25 and 50 sweeps, using a single value of $\rho$ they
took to be an average size. Their distribution was obtained, like
ours, by analysing connected regions of the lattice, and extracting
the size from the size of the region, their criterion being that two
adjacent points belong to the same cluster if the product of $Q(x)$ at
these points is greater than the square of some threshold parameter
introduced because of the algebraic fall-off of the
instanton. However, their methods and results rely heavily on the
assumptions that the instantons are dilute~---~an assumption which we
can see from figs.~\ref{f:cont} and~\ref{f:svsqlink} is an unreliable
one, particularly at their smaller degree of cooling..

Our distributions are peaked
at around $3/m_{0^+}$ at $\beta=2.4$, and at $2/m_{0^+}$ at
$\beta=2.5$.
As in~\cite{smit,chu} we find the smaller scale instantons, those with
$\rho_L\leq 2a$, absent after cooling. Thus at smaller lattice
spacing, we are able to explore the distribution down to smaller
instanton sizes without distortion.

Results presented recently~\cite{peep} showed a plateau in the
distribution of sizes calculated on a lattice with twisted boundary
conditions; our data, calculated using periodic boundary conditions,
show no evidence of such a plateau.

Analagous to the discussion in~\cite{us:dist}, we can deduce a size
distribution of the $SU(2)$ vacuum instantons in the dilute instanton gas
approximation.  In this case the distribution is expected to
be~\cite{dig}
\begin{equation}\label{e:su2dig}
\frac{1}{VS_I}\frac{dS}{d\rho}\sim\rho^{7/3}
\end{equation}
\noindent
Clearly our distributions do not display this behaviour.  It should be
noted that in this case the dilute instanton gas model gives a size
distribution that is infra-red divergent, unlike the ultra-violet
divergence for $O(3)$.

%
%

\section{Instanton--anti-instanton separations}\label{s:seps}
As before, in~\cite{us:dist}, we examined the separations of pairs of
like and unlike objects (\mbox{I-I}, \mbox{A-A} and \mbox{I-A}
pairs). There we found that the average closest separation of unlike
pairs was significantly smaller, with the I-A pairs having a closest
separation approximately 70--75\% that of the I-I or A-A
separations. In section~\ref{s:int} above, we showed how the
interaction between unlike pairs is attractive and so isolated I-A
pairs should be found to occur closer on average than I-I or A-A, but
that I-A pairs closer than a certain minimum distance, given in
eq.~\ref{e:su2rmin}, mutually annihilate, so that we should not expect
to find I-A pairs at smaller separations than when one is located on
the boundary of the other with their sizes such that $R_{\rm
min}=\rho_1=2\rho_2$.  While we did not expect to find this relation
exactly reproduced by the lattice results, as the instantons are not
hyper-spherical and each pair will be influenced by every other object
in the configuration to a greater or lesser degree, we were surprised
to find that it is partially reproduced: the closest separation of
unlike pairs is approximately twice the size of the smaller of the
pair, but is still significantly greater than the size of the
larger. We find that there is no appreciable difference in the ratio
of size in like or unlike pairs, with the larger in the pair being
25--30\% as big again as the smaller. Our lattice data is shown in
table~\ref{t:seps}.

\begin{table}[t]
\begin{center}
\begin{tabular}{ccccccccc}\hline
$\beta$ & $L$ & Method & Sep. & $\langle R_{\rm min}\rangle$ &
\multicolumn{2}{c}{$\langle\rho_{\rm eff}(R_{\rm min})\rangle$} & \#
disc. & \# used\\ \hline\hline
2.4 & 16 & $S(x)$   & I-I & 2.24(15) & 2.93(06) & 3.81(09) & 48 & 213 \\
    &    &          & A-A & 1.89(13) & 2.90(06) & 3.82(08) & 51 & 210 \\
    &    &          & I-A & 6.23(10) & 3.20(06) & 4.17(08) & 46 & 215 \\
    &    &          & U/L & 3.01(26) &          &          &    &     \\
\cline{3-9}
    &    & $|Q(x)|$ & I-I & 2.02(17) & 2.98(05) & 3.67(08) & 56 & 205 \\
    &    &          & A-A & 1.77(13) & 2.95(05) & 3.68(08) & 57 & 204 \\
    &    &          & I-A & 6.52(09) & 3.18(06) & 4.05(06) & 59 & 202 \\
    &    &          & U/L & 3.44(32) &          &          &    &     \\
\hline
2.5 & 24 & $|Q(x)|$ & I-I & 1.62(16) & 2.27(04) & 2.84(07) & 44 & 206 \\
    &    &          & A-A & 1.25(11) & 2.21(04) & 2.83(06) & 29 & 221 \\
    &    &          & I-A & 5.92(11) & 2.50(05) & 3.15(06) & 40 & 210 \\
    &    &          & U/L & 4.14(47) &          &          &    &     \\
\hline
\end{tabular}
\caption{\label{t:seps}The average closest separation of {\em L\/}ike
pairs (I-I and A-A) and {\em U\/}nlike pairs (I-A). `Method' indicates
whether the sizes and locations were calculated using the action
density $S(x)$ or the absolute topological charge density
$|Q(x)|$. $\langle\rho_{\rm eff}(R_{\rm min})\rangle$ gives the
average sizes of the closest objects, averaged over the ensemble of
250 configurations, with the left figure giving the average size of
the smaller object. The last two columns indicate how many
configurations were discounted from the calculations (for example a
configuration with only one instanton was discounted from the I-I
calculations) and how many from the ensemble were used. All sizes in
units of $\xi$.}
\end{center}
\end{table}

Recently~\cite{bubbles} it has been proposed that the instantons and
anti-instantons exist in isolated regions of the lattice, so that
`bubbles' of like objects exist. This could account for the observed
behaviour in the closest separations shown in table~\ref{t:seps}. As a
further probe of this we look at the distribution of separations,
$dN/VdR$, for both like and unlike objects. Our results are given in
figs.~\ref{f:seps1}--\ref{f:seps3}. The double peak structure of
the like separation distribution is evidence that objects are
localised within groups of like objects.

This is a striking result and we should consider whether it could be a
consequence of our instanton finding algorithm. If an isolated
instanton (or anti-instanton) were to have a sufficiently rough peak
structure with several subsidiary peaks, then this would end up
counted more than once.  The consistency between the result obtained
by analysing the $S(x)$ and the $|Q(x)|$ distributions does suggest
that this is not an explanation of the enhanced signal of like species
at small separation.

\begin{figure}[th]
\input{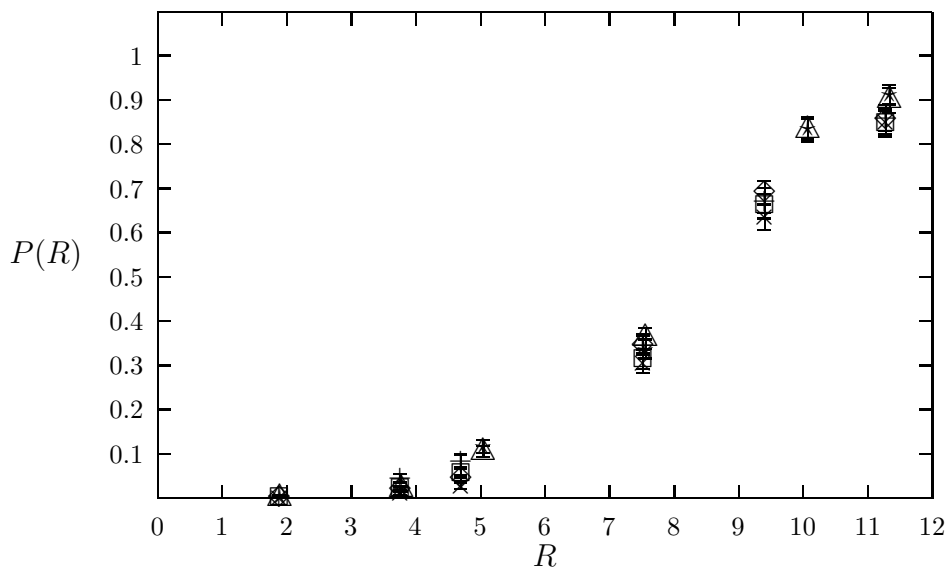}
\caption{\label{f:cent}The proportion, $P(R)$, of objects found within
a distance $R$ (in units of the glueball mass) of the centre-of-mass
calculated as follows:  at
$\beta=2.4, 16^4$ using $S(x)$, I~($\Diamond$), A~($+$), using
$|Q(x)|$, I~($\Box$), A~($\times$); and at $\beta=2.5, 24^4$ using
$|Q(x)|$, I~($\triangle$), A~($\star$).}
\end{figure}

Defining a centre-of-mass $x_c$ for the (anti-)instantons we looked at
the proportion of objects located within certain radial distances of
$x_c$. The results are shown in fig.~\ref{f:cent}. The shape of the
curve, coupled with the data in figs.~\ref{f:seps1}--\ref{f:seps3},
implies that the (anti-)instantons are clustered in many small
bubbles, as one large cluster would give a very sharp rise in the
$P(R)$ curve and a single, broader peak in $dN/VdR$. Clearly this is
an area in which further study is warranted.


%
%

\section{Conclusions}\label{s:su2conc}
Cooling is a local smoothing procedure which enables large scale
excitations to be studied by reducing the quantum fluctuations. In
order to obtain consistent results at different lattice spacings, it
is essential to calibrate the cooling procedure. We explored this
thoroughly in $O(3)$ where we used as a criterion the stability under
cooling of an isolated instanton of a given physical size.  In this
work, we found that for $SU(2)$ (with the Wilson action), the isolated
instanton configurations were extremely stable under cooling. Thus, as
a calibration, we chose instead to create an isolated instanton
solution, add the quantum fluctuations by simulation, and then select
the cooling so that it removed these quantum fluctuations.  Obviously,
it would be possible to cool for some arbitrary amount, as in the
majority of the literature, but we feel there should be some attempt
made to justify the amount of cooling performed, and this calibration
is essential if the cooling is performed, and the results compared, at
different lattice spacings.

Our analysis shows that our resulting cooling prescription should
preserve isolated instantons of size $\rho \ge a$. Of course, cooling
could have a more pronounced effect in the realistic case of dense
instantons: for instance by removing instanton--anti-instanton pairs
which are close together or overlapping. This is hard to calibrate and
must be remembered as a possible source of systematic error.

Compared to previous explorations of topological charge distributions
in que\-n\-ch\-ed $SU(2)$ and $SU(3)$~\cite{polikarpov,chu}, we have been
able to study the topological charge distributions in our cooled
configurations with large statistics, varying lattice spacing and
large physical volume. We have made a comprehensive study of the size
$\rho$ of the peaks in the topological charge distribution.  This size
distribution increases as $\rho^{7/3}$ for a dilute instanton gas. We
find a similar rapid increase for small $\rho$ but then a maximum for
$\rho \approx 0.3$ fm and a rapid decrease at larger $\rho$. Previous
work~\cite{polikarpov} gave evidence for a peak in the size
distribution on cooled configurations, whereas in~cite{chu} there is
indication of a peak only in their more severely cooled data. We have
explored this distribution in more detail and at different lattice
spacings. The absense of important contributions from small objects
agrees with earlier work. In particular, this explains why cooling
gives a reasonable estimate~\cite{pisa} for the topological
susceptibility - since the small objects which would be modified by
cooling are not present anyway. Our conclusions are also compatible
with phenomenological models for the QCD vacuum~\cite{shuryak} which
require instantons of size $\rho \sim {1 \over 3}$fm with a density
$n\sim 1$fm$^-4$.

Our results show some evidence that the peak in the size distribution
is at smaller $\rho$ for the smaller lattice spacing ($\beta=2.5$).
We also find a slight difference in the instanton density, namely
$n(\beta=2.4)=0.71$ fm$^-4$ and $n(\beta=2.5)=0.96$ fm$^-4$.  This
residual dependence on $\beta$ might be caused by our cooling
calibration being not quite optimum and so corresponding to slightly
stronger cooling at $\beta=2.4$. Another possible explanation is that
lattice artefacts in the instanton solution are important for $\rho <
2a$ and so are more significant at the larger lattice spacing.  This
uncertainty can be resolved by varying the cooling amount and by
reducing the lattice spacing even further.

We have studied the relative disposition of instantons (I) and
anti-instantons (A).  We obtain the expected result that there is a
minimum IA separation~---~since they will tend to annihilate if too
close. We find, however, a surprising result: that there is a signal
from II and AA pairs at close separation. This could arise if
instantons were to exist in localised groups, well-separated from
localised groups of anti-instantons.  This clearly needs further work
to establish the details of such a mechanism.

\subsection*{Acknowledgements}
We are indebted to Claus Montonen for bringing our attention to
references~\cite{forster} and~\cite{palmer}. This research was carried
out as part of the EC Programme ``Human Capital and
Mobility''~---~project number {\sc erb--chrx--ct92--0051}.

%
%
\newpage
\section*{Figures}


\begin{figure}[th]
\input{sq16.dist.tex}
\caption{\label{f:dist1}Size distributions ($\rho$ in
units of the glueball mass) calculated on a $16^4$ lattice at
$\beta=2.4$.}
\end{figure}
\begin{figure}[bh]
\setlength{\unitlength}{0.240900pt}
\ifx\plotpoint\undefined\newsavebox{\plotpoint}\fi
\begin{picture}(1500,900)(0,0)
\font\gnuplot=cmr10 at 10pt
\gnuplot
\sbox{\plotpoint}{\rule[-0.200pt]{0.400pt}{0.400pt}}%
\put(220.0,113.0){\rule[-0.200pt]{292.934pt}{0.400pt}}
\put(220.0,113.0){\rule[-0.200pt]{0.400pt}{184.048pt}}
\put(220.0,113.0){\rule[-0.200pt]{4.818pt}{0.400pt}}
\put(198,113){\makebox(0,0)[r]{0}}
\put(1416.0,113.0){\rule[-0.200pt]{4.818pt}{0.400pt}}
\put(220.0,189.0){\rule[-0.200pt]{4.818pt}{0.400pt}}
\put(198,189){\makebox(0,0)[r]{0.2}}
\put(1416.0,189.0){\rule[-0.200pt]{4.818pt}{0.400pt}}
\put(220.0,266.0){\rule[-0.200pt]{4.818pt}{0.400pt}}
\put(198,266){\makebox(0,0)[r]{0.4}}
\put(1416.0,266.0){\rule[-0.200pt]{4.818pt}{0.400pt}}
\put(220.0,342.0){\rule[-0.200pt]{4.818pt}{0.400pt}}
\put(198,342){\makebox(0,0)[r]{0.6}}
\put(1416.0,342.0){\rule[-0.200pt]{4.818pt}{0.400pt}}
\put(220.0,419.0){\rule[-0.200pt]{4.818pt}{0.400pt}}
\put(198,419){\makebox(0,0)[r]{0.8}}
\put(1416.0,419.0){\rule[-0.200pt]{4.818pt}{0.400pt}}
\put(220.0,495.0){\rule[-0.200pt]{4.818pt}{0.400pt}}
\put(198,495){\makebox(0,0)[r]{1}}
\put(1416.0,495.0){\rule[-0.200pt]{4.818pt}{0.400pt}}
\put(220.0,571.0){\rule[-0.200pt]{4.818pt}{0.400pt}}
\put(198,571){\makebox(0,0)[r]{1.2}}
\put(1416.0,571.0){\rule[-0.200pt]{4.818pt}{0.400pt}}
\put(220.0,648.0){\rule[-0.200pt]{4.818pt}{0.400pt}}
\put(198,648){\makebox(0,0)[r]{1.4}}
\put(1416.0,648.0){\rule[-0.200pt]{4.818pt}{0.400pt}}
\put(220.0,724.0){\rule[-0.200pt]{4.818pt}{0.400pt}}
\put(198,724){\makebox(0,0)[r]{1.6}}
\put(1416.0,724.0){\rule[-0.200pt]{4.818pt}{0.400pt}}
\put(220.0,801.0){\rule[-0.200pt]{4.818pt}{0.400pt}}
\put(198,801){\makebox(0,0)[r]{1.8}}
\put(1416.0,801.0){\rule[-0.200pt]{4.818pt}{0.400pt}}
\put(220.0,877.0){\rule[-0.200pt]{4.818pt}{0.400pt}}
\put(198,877){\makebox(0,0)[r]{2}}
\put(1416.0,877.0){\rule[-0.200pt]{4.818pt}{0.400pt}}
\put(220.0,113.0){\rule[-0.200pt]{0.400pt}{4.818pt}}
\put(220,68){\makebox(0,0){0}}
\put(220.0,857.0){\rule[-0.200pt]{0.400pt}{4.818pt}}
\put(372.0,113.0){\rule[-0.200pt]{0.400pt}{4.818pt}}
\put(372,68){\makebox(0,0){1}}
\put(372.0,857.0){\rule[-0.200pt]{0.400pt}{4.818pt}}
\put(524.0,113.0){\rule[-0.200pt]{0.400pt}{4.818pt}}
\put(524,68){\makebox(0,0){2}}
\put(524.0,857.0){\rule[-0.200pt]{0.400pt}{4.818pt}}
\put(676.0,113.0){\rule[-0.200pt]{0.400pt}{4.818pt}}
\put(676,68){\makebox(0,0){3}}
\put(676.0,857.0){\rule[-0.200pt]{0.400pt}{4.818pt}}
\put(828.0,113.0){\rule[-0.200pt]{0.400pt}{4.818pt}}
\put(828,68){\makebox(0,0){4}}
\put(828.0,857.0){\rule[-0.200pt]{0.400pt}{4.818pt}}
\put(980.0,113.0){\rule[-0.200pt]{0.400pt}{4.818pt}}
\put(980,68){\makebox(0,0){5}}
\put(980.0,857.0){\rule[-0.200pt]{0.400pt}{4.818pt}}
\put(1132.0,113.0){\rule[-0.200pt]{0.400pt}{4.818pt}}
\put(1132,68){\makebox(0,0){6}}
\put(1132.0,857.0){\rule[-0.200pt]{0.400pt}{4.818pt}}
\put(1284.0,113.0){\rule[-0.200pt]{0.400pt}{4.818pt}}
\put(1284,68){\makebox(0,0){7}}
\put(1284.0,857.0){\rule[-0.200pt]{0.400pt}{4.818pt}}
\put(1436.0,113.0){\rule[-0.200pt]{0.400pt}{4.818pt}}
\put(1436,68){\makebox(0,0){8}}
\put(1436.0,857.0){\rule[-0.200pt]{0.400pt}{4.818pt}}
\put(220.0,113.0){\rule[-0.200pt]{292.934pt}{0.400pt}}
\put(1436.0,113.0){\rule[-0.200pt]{0.400pt}{184.048pt}}
\put(220.0,877.0){\rule[-0.200pt]{292.934pt}{0.400pt}}
\put(-15,495){\makebox(0,0){$dN/Vd\rho\times10^4$}}
\put(828,23){\makebox(0,0){$\rho$}}
\put(220.0,113.0){\rule[-0.200pt]{0.400pt}{184.048pt}}
\put(1306,812){\makebox(0,0)[r]{$|Q(x)|$ on $24^4, \beta=2.5$}}
\put(1350,812){\raisebox{-.8pt}{\makebox(0,0){$\Box$}}}
\put(298,113){\raisebox{-.8pt}{\makebox(0,0){$\Box$}}}
\put(377,196){\raisebox{-.8pt}{\makebox(0,0){$\Box$}}}
\put(454,650){\raisebox{-.8pt}{\makebox(0,0){$\Box$}}}
\put(533,801){\raisebox{-.8pt}{\makebox(0,0){$\Box$}}}
\put(611,643){\raisebox{-.8pt}{\makebox(0,0){$\Box$}}}
\put(688,555){\raisebox{-.8pt}{\makebox(0,0){$\Box$}}}
\put(767,358){\raisebox{-.8pt}{\makebox(0,0){$\Box$}}}
\put(845,231){\raisebox{-.8pt}{\makebox(0,0){$\Box$}}}
\put(924,175){\raisebox{-.8pt}{\makebox(0,0){$\Box$}}}
\put(1001,132){\raisebox{-.8pt}{\makebox(0,0){$\Box$}}}
\put(1080,120){\raisebox{-.8pt}{\makebox(0,0){$\Box$}}}
\put(1158,119){\raisebox{-.8pt}{\makebox(0,0){$\Box$}}}
\put(1235,115){\raisebox{-.8pt}{\makebox(0,0){$\Box$}}}
\put(1314,113){\raisebox{-.8pt}{\makebox(0,0){$\Box$}}}
\put(1392,113){\raisebox{-.8pt}{\makebox(0,0){$\Box$}}}
\put(1328.0,812.0){\rule[-0.200pt]{15.899pt}{0.400pt}}
\put(1328.0,802.0){\rule[-0.200pt]{0.400pt}{4.818pt}}
\put(1394.0,802.0){\rule[-0.200pt]{0.400pt}{4.818pt}}
\put(220,113){\usebox{\plotpoint}}
\put(210.0,113.0){\rule[-0.200pt]{4.818pt}{0.400pt}}
\put(210.0,113.0){\rule[-0.200pt]{4.818pt}{0.400pt}}
\put(298,113){\usebox{\plotpoint}}
\put(288.0,113.0){\rule[-0.200pt]{4.818pt}{0.400pt}}
\put(288.0,113.0){\rule[-0.200pt]{4.818pt}{0.400pt}}
\put(377.0,181.0){\rule[-0.200pt]{0.400pt}{7.227pt}}
\put(367.0,181.0){\rule[-0.200pt]{4.818pt}{0.400pt}}
\put(367.0,211.0){\rule[-0.200pt]{4.818pt}{0.400pt}}
\put(454.0,609.0){\rule[-0.200pt]{0.400pt}{19.513pt}}
\put(444.0,609.0){\rule[-0.200pt]{4.818pt}{0.400pt}}
\put(444.0,690.0){\rule[-0.200pt]{4.818pt}{0.400pt}}
\put(533.0,767.0){\rule[-0.200pt]{0.400pt}{16.140pt}}
\put(523.0,767.0){\rule[-0.200pt]{4.818pt}{0.400pt}}
\put(523.0,834.0){\rule[-0.200pt]{4.818pt}{0.400pt}}
\put(611.0,621.0){\rule[-0.200pt]{0.400pt}{10.600pt}}
\put(601.0,621.0){\rule[-0.200pt]{4.818pt}{0.400pt}}
\put(601.0,665.0){\rule[-0.200pt]{4.818pt}{0.400pt}}
\put(688.0,530.0){\rule[-0.200pt]{0.400pt}{12.045pt}}
\put(678.0,530.0){\rule[-0.200pt]{4.818pt}{0.400pt}}
\put(678.0,580.0){\rule[-0.200pt]{4.818pt}{0.400pt}}
\put(767.0,337.0){\rule[-0.200pt]{0.400pt}{9.877pt}}
\put(757.0,337.0){\rule[-0.200pt]{4.818pt}{0.400pt}}
\put(757.0,378.0){\rule[-0.200pt]{4.818pt}{0.400pt}}
\put(845.0,217.0){\rule[-0.200pt]{0.400pt}{6.504pt}}
\put(835.0,217.0){\rule[-0.200pt]{4.818pt}{0.400pt}}
\put(835.0,244.0){\rule[-0.200pt]{4.818pt}{0.400pt}}
\put(924.0,164.0){\rule[-0.200pt]{0.400pt}{5.059pt}}
\put(914.0,164.0){\rule[-0.200pt]{4.818pt}{0.400pt}}
\put(914.0,185.0){\rule[-0.200pt]{4.818pt}{0.400pt}}
\put(1001.0,128.0){\rule[-0.200pt]{0.400pt}{2.168pt}}
\put(991.0,128.0){\rule[-0.200pt]{4.818pt}{0.400pt}}
\put(991.0,137.0){\rule[-0.200pt]{4.818pt}{0.400pt}}
\put(1080.0,118.0){\rule[-0.200pt]{0.400pt}{0.964pt}}
\put(1070.0,118.0){\rule[-0.200pt]{4.818pt}{0.400pt}}
\put(1070.0,122.0){\rule[-0.200pt]{4.818pt}{0.400pt}}
\put(1158.0,117.0){\rule[-0.200pt]{0.400pt}{0.964pt}}
\put(1148.0,117.0){\rule[-0.200pt]{4.818pt}{0.400pt}}
\put(1148.0,121.0){\rule[-0.200pt]{4.818pt}{0.400pt}}
\put(1235.0,114.0){\rule[-0.200pt]{0.400pt}{0.482pt}}
\put(1225.0,114.0){\rule[-0.200pt]{4.818pt}{0.400pt}}
\put(1225.0,116.0){\rule[-0.200pt]{4.818pt}{0.400pt}}
\put(1314,113){\usebox{\plotpoint}}
\put(1304.0,113.0){\rule[-0.200pt]{4.818pt}{0.400pt}}
\put(1304.0,113.0){\rule[-0.200pt]{4.818pt}{0.400pt}}
\put(1392,113){\usebox{\plotpoint}}
\put(1382.0,113.0){\rule[-0.200pt]{4.818pt}{0.400pt}}
\put(1382.0,113.0){\rule[-0.200pt]{4.818pt}{0.400pt}}
\end{picture}
\caption{\label{f:dist2}Size distributions ($\rho$ in
units of the glueball mass) calculated on a $24^4$ lattice at
$\beta=2.5$.}
\end{figure}
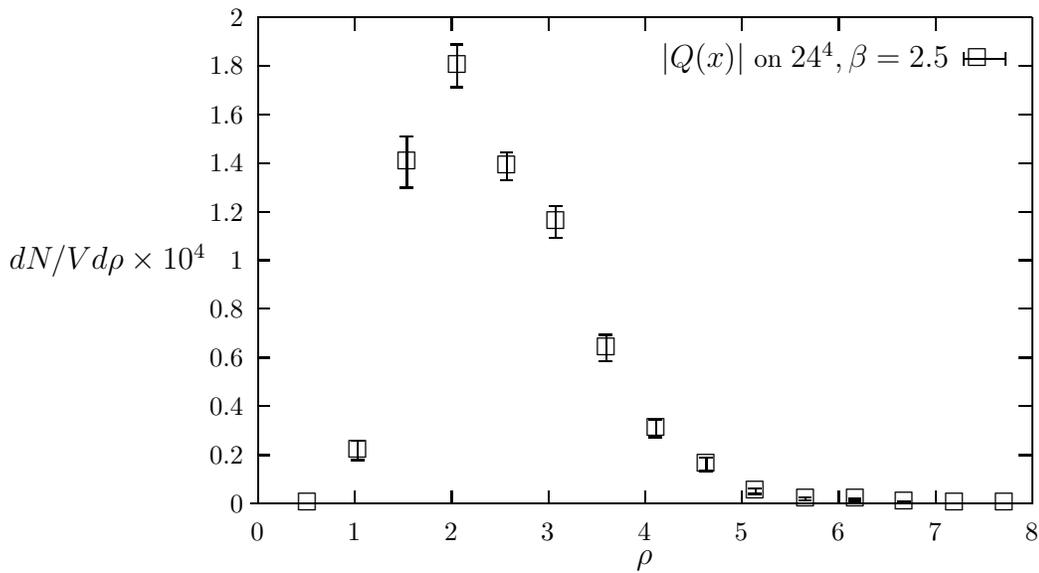

\begin{figure}[th]
\input{s16.seps.tex}
\caption{\label{f:seps1}The data for the distribution of separations
of like and unlike pairs (in units of the glueball mass),
calculated at $\beta=2.4$ on a $16^4$ lattice using $S(x)$.}
\end{figure}
\begin{figure}[bh]
\input{q16.seps.tex}
\caption{\label{f:seps2}The data for the distribution of separations
of like and unlike pairs (in units of the glueball mass),
calculated at $\beta=2.4$ on a $16^4$ lattice using $|Q(x)|$.}
\end{figure}
\begin{figure}[bh]
\input{q24.seps.tex}
\caption{\label{f:seps3}The data for the distribution of separations
of like and unlike pairs (in units of the glueball mass),
calculated at $\beta=2.5$ on a $24^4$ lattice using $|Q(x)|$.}
\end{figure}

\clearpage

%
%
%


\begin{thebibliography}{99}
\bibitem{tepersu2} M.~Teper, Phys. Lett. {\bf 162B}, 357 (1985).
\bibitem{1stinst} E.-M.~Ilgenfritz, M.L.~Laursen,
M.~M\"uller-Preussker, G.~Schierholz, H.~Schiller, Nucl. Phys. {\bf
B268}, 693 (1986).
\bibitem{polikarpov} M.I.~Polikarpov, A.I.~Veselov, Nucl. Phys. {\bf
B297}, 34 (1988).
\bibitem{belavin} A.A.~Belavin, A.M.~Polyakov, A.S.~Schwatrz and
Yu.S.~Tyupkin, Phys. Lett. {\bf 59B}, 85 (1975).
\bibitem{forster} D.~F\"orster, Phys. Lett. {\bf 66B}, 279 (1977).
\bibitem{palmer} W.~Palmer \& S.S.~Pinsky, Phys. Rev. {\bf D21}, 551
(1980).
\bibitem{us:lat94} {\em Instanton size distributions from calibrated
cooling\/}, C.~Michael \& P.S.~Spencer, Liverpool Preprint LTH-337,
Helsinki preprint HU-TFT-94-45, to appear in the proceedings of
LATTICE'94.
\bibitem{chengli} {\em Gauge theory of elementary particle physics\/},
Ta-Pei~Cheng and Ling-Fong~Li, Oxford University Press (1984).
\bibitem{smit} M.L.~Laursen, J.~Smit, J.C.~Vink, Nucl. Phys. {\bf
B343}, 522 (1990).
\bibitem{us:cool} {\em Topography of the cooled $O(3)$
vacuum\/}, P.S.~Spencer and C.~Michael, Liverpool preprint LTH 328,
hep-lat/9401011, submitted to J.~Phys.~G.
\bibitem{us:dist} C.~Michael \& P.S.~Spencer, Phys. Rev.~{\bf D50},
7570 (1994).
\bibitem{mjtinst} {\em Large Instantons In Lattice Gauge
Theory And Their Stability Under ``Cooling''\/}, M~.Teper, Oxford
Univerity preprint~59/88
\bibitem{green} A.M.~Green, C.~Michael, J.E.~Paton, M.E.~Sainio,
Int. J.~Mod. Phys. {\bf E2}, 479 (1993).
\bibitem{karsch} J.~Fingberg, U.M.~Heller, F.~Karsch Nucl.~Phys.~{\bf
B392}, 493 (1993).
\bibitem{campopl} M.~Campostrini {\em et al\/}, Nucl. Phys. {\bf
B329}, 683 (1990).
\bibitem{grandyl94} {\em Topological density and instantons on a
lattice\/}, J.~Grandy \& R.~Gupta, contribution to LATTICE~'94.
\bibitem{su2mass} C.~Michael, G.A.~Tickle, M.~Teper Phys. Lett. {\bf
207B}, 313 (1988).
\bibitem{chu} M.-C.~Chu, J.M.~Grandy, S.~Huang, J.W.~Negele,
Nucl. Phys.~B (Proc. Suppl.)~{\bf 34}, 170, (1994);\\ M.-C.~Chu,
J.M.~Grandy, S.~Huang, J.W.~Negele, Phys. Rev.~{\bf D49}, 6039 (1994).
\bibitem{peep} {\em Peeping into the $SU(2)$ gauge vacuum\/},
A.~Gonzales-Arroyo, FTUAM-94-26, hep-lat/9502014, contribution to
LATTICE~'94.
\bibitem{dig} This relation occurs in many papers and
textbooks. Amongst others:\\
{\em Gauge fields and strings\/} A.M.~Polyakov, North Holland
(1982);\\
{\em The uses of instantons\/}, S.~Coleman, in {\em The Whys of
Subnuclear Physics\/}, \indent ed.~A.~Zichichi, Plenum Press, New York
(1979);\\
{\em Solitons and instantons\/}, R.~Rajaraman, North Holland (1982).
\bibitem{bubbles} A.~Gonzales-Arroyo, talk given at Cortona.
\bibitem{shuryak} E.~Shuryak, Rev. Mod. Phys. {\bf 65}, 1 (1993).
\bibitem{pisa} B.~Alles {\em et al}, Phys. Rev.~{\bf D48}, 2284
(1993).

\end{thebibliography}
\end{document}